\documentclass{aa}
\usepackage{graphicx}
\usepackage{txfonts}
 \begin{document}

   \title{Stability of toroidal magnetic fields
          in rotating stellar radiation zones}

   \author{L.\,L.~Kitchatinov\inst{1,2} \and G.~R\"udiger\inst{1}
          }

   \offprints{G.~R\"udiger}

   \institute{Astrophysikalisches Institut Potsdam, An der Sternwarte 16,
              D-14482, Potsdam, Germany \\
              \email{lkitchatinov@aip.de; gruediger@aip.de}
         \and
             Institute for Solar-Terrestrial Physics, PO Box
             4026, Irkutsk 664033, Russia \\
             \email{kit@iszf.irk.ru}
             }

   \date{Received ; accepted }
  \abstract
  {}
  {The questions of how strong magnetic fields can be stored in
rotating stellar radiative zones without being subjected to
pinch-type instabilities and how much radial mixing is produced if
the fields are unstable are addressed.}
  {Linear equations are derived for weak disturbances of magnetic and
velocity fields which are global in horizontal dimensions but
short--scaled in radius. The equations are solved to evaluate the
stability of toroidal field patterns with one or two latitudinal
belts under the influence of a rigid basic rotation. Hydrodynamic
stability of latitudinal differential rotation is also
considered.}
  {The magnetic instability is essentially three--dimensional.
It does not exist in a 2D formulation with strictly horizontal
disturbances on decoupled spherical shells. Only stable
(magnetically modified) $r$-modes are found in this case. The
instability recovers in 3D. The most rapidly growing modes for the
Sun have radial scales smaller than 1 Mm. The finite thermal
conductivity makes a strong destabilizing effect. The marginal
field strength for the onset of the instability in the upper part
of the solar radiative zone is about 600~G. The toroidal field can
only slightly exceed this critical value for otherwise the radial
mixing produced by the instability would be too strong to be
compatible with the observed lithium abundance. Also the threshold
for hydrodynamic instability of differential rotation which exists
in 2D is lowered in 3D. When radial displacements are included,
the value of 28\% for critical shear is reduced to 21\%.
  }
  {}


   \titlerunning{Magnetic instabilities in stellar radiation zones}

   \authorrunning{L.\,L.~Kitchatinov \& G.~R\"udiger}

   \maketitle
\section{Introduction}
The old problem of hydromagnetic stability of stellar radiative
zones attains a renewed interest in relation with Ap stars
magnetism (Braithwaite \& Spruit \cite{BS04}), the solar
tachocline (Gilman \cite{G05}) and transport processes in stably
stratified stellar interiors (Barnes et al. \cite{BCM99}).

Which fields the stellar radiative cores can possess is rather
uncertain. The resistive decay in radiation zones is so slow that
primordial fields can be stored there (Cowling \cite{C45}).
Whether the fields of 10$^5$~G which can influence g-modes of
solar oscillations (Rashba et al. \cite{Rea06})  or still stronger
fields which can cause neutrino oscillations (Burgess et al.
\cite{Bea03}) can indeed survive inside the Sun mainly depends on
their stability.

Among several instabilities to which the fields can be subjected
(Acheson \cite{A78}), the current-driven (pinch-type) instability
of toroidal fields (Tayler \cite{T73}) is probably the most
relevant one because it proceeds via almost horizontal
displacements. The radial motions in radiative cores are strongly
suppressed by buoyancy. Watson (\cite{W81}) estimated the ratio of
radial ($u_r$) to horizontal ($u_\theta$) velocities of slow
(subsonic) motions in rotating stars as
 $
   u_r/u_\theta \sim \Omega^2/N^2 ,
 $
where $\Omega$ is the basic angular velocity and $N$ the buoyancy
(Brunt-V\"ais\"al\"a) frequency. This ratio is small in radiative
cores (Fig.~\ref{f1}). If the radial velocities are completely
neglected, the stability analysis can be done in 2D approximation
with decoupled spherical shells (Watson \cite{W81}; Gilman \& Fox
\cite{GF97}). There is, however, some radial motion still excited.
How much radial mixing the Tayler (\cite{T73}) instability
produces is not well-known so far. As the radial mixing is
relevant to the transport of light elements, a theory  of  the
mixing compared with the observed abundances can help to restrict
the amplitudes of the internal magnetic fields (Barnes et al.
\cite{BCM99}). In the present paper the vertical mixing produced
by the Tayler instability is estimated and then used to evaluate
the upper limit on the magnetic field amplitude.

In this paper, the equations governing linear stability of
toroidal magnetic fields in differentially rotating radiation zone
are derived.  They are solved for two latitudinal profiles of the
toroidal field with one and with two belts in latitude but only
for rigid rotation. The unstable modes are expected to have the
longest possible horizontal scales (Spruit \cite{S99}).
Accordingly, our equations are global in horizontal dimensions.
They are, however, local in radius, i.e. the radial scales are
assumed short, $kr \gg 1$ ($k$ is radial wave number). The
computations confirm that the most rapidly growing modes have $kr
\sim 10^3$ but they are global in latitude. The derived equations
reproduce the 2D approximation as a special limit. It is shown
with an exactly solvable case that for rigid rotation the Tayler
instability is missing in the 2D approximation. It recovers,
however, in 3D case. Finite diffusion is found important for the
instability. The minimum field producing the instability is
strongly reduced by allowance for finite thermal conductivity.
This field amplitude is about 600~G for the upper part of the
solar radiation zone. Considering the light elements transport by
the Tayler instability we find that the field strength can be only
slightly above the marginal value. Otherwise, the mixing would not
be compatible with the observed lithium abundance.
\section{The model}
\subsection{Background state and basic assumptions}
The stability of rotating radiation zone of a star containing
magnetic field is considered. The field is assumed axisymmetric
and purely toroidal. Hence, it can be written as
\begin{equation}
   {\vec B} = {\vec e}_\phi r\sin\theta\sqrt{\mu_0\rho}\
   \Omega_\mathrm{A}\left( r,\theta\right)
   \label{1}
\end{equation}
in terms of the Alfv\'en angular frequency $\Omega_\mathrm{A}$. In
this equation, $\rho$ is the density, $r,\theta ,\phi$ are the
usual spherical coordinates and ${\vec e}_\phi$ it the azimuthal
unit vector. Equation (\ref{1}) automatically ensures that the
toroidal field component vanishes -- as it must -- at the rotation
axis. Centrifugal and magnetic forces are assumed small compared
to gravity, $g/r \gg \Omega^2$ and  $g/r \gg \Omega^2_\mathrm{A}$.
Deviation of the fluid stratification from spherical symmetry can
thus be neglected.

\begin{figure}
   \centering
   \includegraphics[width=8.8cm, height=6.0cm]{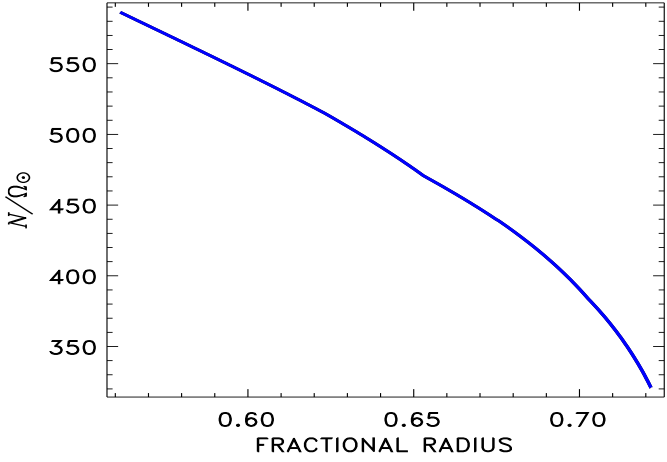}
   \caption{The buoyancy frequency (\ref{2})  in the upper part
            of the radiative core of the Sun after the model by
            Stix \& Skaley (\cite{SS90}).
            The convection zone of the model includes the overshoot layer
            so that $N/\Omega_\odot$ is large immediately beneath
            the convection zone.
              }
   \label{f1}
\end{figure}

The stabilizing effect of a subadiabatic stratification of the
radiative core is characterized by the buoyancy frequency,
\begin{equation}
   N^2 = \frac{g}{C_\mathrm{p}}\frac{\partial s}{\partial r} ,
   \label{2}
\end{equation}
where $s = C_\mathrm{v}\ln \left( P/\rho^\gamma\right)$ is the
specific entropy of ideal gas. The frequency $N$ is very large  in
the radiative cores of not too rapidly rotating stars like the Sun
($N \gg \Omega$, $N \gg \Omega_\mathrm{A}$,  see Fig.~\ref{f1}).

The larger is $N$ the more the radial displacements are opposed by
the buoyancy force. Radial velocities should therefore be small.
They are often neglected in stability analysis what might be
dangerous as certain instabilities may even be suppressed by the
neglect (we shall see later how this indeed happens for the Tayler
instability).

The consequences of the neglect of the radial velocity
perturbations, $u'_r$, can be seen from the expression for
(divergence-free) velocity, ${\vec u}'$, in spherical geometry in
terms of the two scalar potentials for the poloidal,
$P_\mathrm{v}$, and toroidal, $T_\mathrm{v}$, flows
 \begin{eqnarray}
  {\vec u}' &=& \frac{{\vec e}_r}{r^2}\hat{\cal L}P_\mathrm{v}
  - \frac{{\vec e}_\theta}{r}\left(\frac{1}{\sin\theta}
  \frac{\partial T_\mathrm{v}}{\partial\phi} +
  \frac{\partial^2 P_\mathrm{v}}{\partial r\partial\theta}\right)+
  \nonumber \\
  &+&\frac{{\vec e}_\phi}{r}\left(\frac{\partial
  T_\mathrm{v}}{\partial\theta} -
  \frac{1}{\sin\theta}\frac{\partial^2 P_\mathrm{v}}
  {\partial r\partial\phi}\right),
  \nonumber \\
  \hat{\cal L} &=& \frac{1}{\sin\theta}\frac{\partial}{\partial\theta}
  \sin\theta\frac{\partial}{\partial\theta} +
  \frac{1}{\sin^2\theta}\frac{\partial^2}{\partial\phi^2}
  \label{3}
\end{eqnarray}
(cf. Chandrasekhar \cite{C61}). Here and in the following
disturbances are marked by dashes. For zero radial velocity, the
whole class of disturbances corresponding to the poloidal flow
vanishes. Then the remaining toroidal flow can produce only
toroidal magnetic disturbances so that in the expression for the
magnetic field perturbations (${\vec B}'$),
\begin{eqnarray}
  {\vec B}' &=& \frac{{\vec e}_r}{r^2}\hat{\cal L}P_\mathrm{m}
  - \frac{{\vec e}_\theta}{r}\left(\frac{1}{\sin\theta}
  \frac{\partial T_\mathrm{m}}{\partial\phi} +
  \frac{\partial^2 P_\mathrm{m}}{\partial r\partial\theta}\right)+
  \nonumber \\
  &+& \frac{{\vec e}_\phi}{r}\left(\frac{\partial
  T_\mathrm{m}}{\partial\theta} -
  \frac{1}{\sin\theta}\frac{\partial^2 P_\mathrm{m}}
  {\partial r\partial\phi}\right) ,
  \label{4}
\end{eqnarray}
the poloidal field potential becomes zero, $P_\mathrm{m} = 0$.
Reducing the class of disturbances may switch-off some
instabilities.

It can be seen from Eq.~(\ref{3}) that the horizontal part of the
poloidal flow can remain unchanged when the radial velocity is
reduced and the radial {\em scale} of the flow is reduced
proportionally. The disturbances can thus avoid the stabilizing
effect of the stratification by decreasing their radial scale.

Our stability analysis is local in the radial dimension, i.e. we
use Fourier modes $\exp (\mathrm{i}kr)$ with $kr \gg 1$. It will
be confirmed that the most unstable modes do indeed prefer short
radial scales. The analysis remains, however, global in horizontal
dimensions.

Instabilities of toroidal fields or differential rotation proceed
via not compressive disturbances. Characteristic growth rates of
the instabilities are small compared to p-modes frequencies. In
the short-wave approximation the velocity field can be assumed
divergence-free, $\mathrm{div} {\vec u}' = 0$. Note that even a
slow motion in a stratified fluid can be divergent if its spatial
scale in radial direction is not small compared to scale height.
In the short-wave approximation (in radius) the divergency can,
however, be neglected.

Our next assumption concerns the pressure. More precisely, local
thermal disturbances occur at constant pressure so that
$\rho'/\rho = -T'/T$ or $s' = -C_\mathrm{p}\rho'/\rho$. This
assumption is again justified by the incompressible nature of the
perturbations. Another interpretation of this assumption is given
by Acheson (\cite{A78}). Acheson assumed zero disturbances of
total (including magnetic) pressure to involve magnetic buoyancy
instability. In our derivations, assumptions on constant total or
only gas pressure are identical because the effect of magnetic
buoyancy appears in higher order in $(kr)^{-1}$ compared to the
terms kept in the equations of the next section.
\subsection{Equations}\label{equations}
We start from the linearized equations for small velocity perturbations, i.e.
\begin{eqnarray}
 \frac{\partial{\vec u}'}{\partial t}
 &+& \left({\vec u}\cdot\nabla\right){\vec u}'
 + \left({\vec u}'\cdot\nabla\right){\vec u}
 + \frac{1}{{\mu_0\rho}}\left(\nabla\left({\vec B}\cdot{\vec B}'\right)\right.
 \nonumber \\
 &-& \left.\left({\vec B}\cdot\nabla\right){\vec B}'
 - \left({\vec B}'\cdot\nabla\right){\vec B}\right)
 =-\left(\frac{1}{\rho}\nabla P\right)' + \nu\Delta{\vec u}'
  \label{5}
\end{eqnarray}
magnetic field,
\begin{equation}
   \frac{\partial{\vec B}'}{\partial t} =
   \nabla\times\left( {\vec u}\times{\vec B}'
   + {\vec u}'\times{\vec B} - \eta\nabla\times{\vec B}'\right) ,
   \label{6}
\end{equation}
and entropy,
\begin{equation}
  \frac{\partial s'}{\partial t} + {\vec u}\cdot \nabla s' +
  {\vec u}'\cdot\nabla s = \frac{C_\mathrm{p}\chi}{T}\Delta T' .
  \label{7}
\end{equation}
The basic flow is a rotation with nonuniform angular velocity,
$\Omega= \Omega (r,\theta)$, and the mean magnetic field is the
toroidal one of the form   (\ref{1}). Perturbations of velocity
and magnetic field are expressed in terms of scalar potentials
after  (\ref{3}) and (\ref{4}). The identities
\begin{eqnarray}
   r \left({\vec r}\cdot\nabla\times{\vec B}'\right) &=&
   \hat{\cal L} T_\mathrm{m},\ \ \
   r \left({\vec r}\cdot{\vec B}'\right) = \hat{\cal L}P_\mathrm{m},
   \nonumber \\
   r \left({\vec r}\cdot\nabla\times{\vec u}'\right) &=&
   \hat{\cal L} T_\mathrm{v},\ \ \
   r^3\left({\vec r}\cdot\nabla\times\nabla\times{\vec u}'\right) =
   \left(\hat{\cal{L}} + r^2\frac{\partial^2}{\partial r}\right)
   \hat{\cal L}P_\mathrm{v}
   \nonumber
\end{eqnarray}
are used to reformulate the equations in terms of the potentials.
The radial component of Eq.~(\ref{6}) then gives the equation for
the poloidal magnetic field and the  radial components of {\em
curled} equations (\ref{5}) and (\ref{6}) give the equations for
toroidal flow and toroidal magnetic field,  respectively\footnote{
the radial component of Eq.~(\ref{5}) curled twice provides the
equation for the poloidal flow}.

The perturbations are considered as Fourier modes in time, azimuth
and radius in the form of $\mathrm{exp}\left(\mathrm{i}(-\omega t
+ m\phi + kr)\right)$. For an instability, the eigenvalue $\omega$
should possess a positive imaginary part. Only the highest-order
terms in $kr$ for the same variable were kept in the same
equation.

When deriving the poloidal flow equation, the pressure term was
transformed as follows
\begin{eqnarray}
   {\vec r}\cdot\nabla\times\nabla\times\left(\frac{1}{\rho}
   \nabla P\right)' &=&
   -{\vec r}\cdot\nabla\times\left(\frac{1}{\rho^2}
   (\nabla\rho)\times(\nabla P)\right)' =
   \nonumber \\
   \frac{\vec r}{C_\mathrm{p}}\cdot\nabla\times\left(\frac{1}{\rho}
   (\nabla s)\times(\nabla P)\right)' &=&
   -\frac{\vec r}{C_\mathrm{p}}\cdot\nabla\times
   \left({\vec g}\times\nabla s'\right) =  \frac{g}{rC_\mathrm{p}}
   \hat{\cal L}s' .
   \nonumber
\end{eqnarray}

In order to use normalized variables the time is measured in units
of $\Omega_0^{-1}$ ($\Omega_0$ is a characteristic angular
velocity), the  velocities are scaled in units of $r\Omega_0$, the
normalized  frequency ($\hat\omega$) is measured in units of
$\Omega_0$, and the other normalized variables are
\begin{eqnarray}
   A &=& \frac{k}{\Omega_0 r^2\sqrt{\mu_0\rho}}\ P_\mathrm{m},\ \
   B = \frac{1}{\Omega_0 r^2\sqrt{\mu_0\rho}}\ T_\mathrm{m},\ \
   V = \frac{k}{\Omega_0 r^2}\ P_\mathrm{v},\ \
   \nonumber \\
   W &=& \frac{1}{\Omega_0 r^2}\ T_\mathrm{v},\ \
   S = \frac{\mathrm{i} k r g}{C_\mathrm{p} r N^2}\ s' ,\ \ \ \
   \hat\Omega = \frac{\Omega}{\Omega_0},\ \ \ \
   \hat\Omega_\mathrm{A} = \frac{\Omega_\mathrm{A}}{\Omega_0} .
   \label{10}
\end{eqnarray}
Introducing the factors $kr$ in the normalizations of poloidal
potentials makes them equal by order of magnitude to the  toroidal
potentials, while in Eqs. (\ref{3}) and (\ref{4}) it was
$P_\mathrm{v} \ll rT_\mathrm{v}$, $P_\mathrm{m} \ll
rT_\mathrm{m}$.

The equation for the poloidal flow then reads
\begin{eqnarray}
   \hat\omega\left(\hat{\cal L}V\right) &=&
   -\hat{\lambda}^2\left(\hat{\cal L}S\right)
   - \mathrm{i}\frac{\epsilon_\nu}{\hat{\lambda}^2}\left(\hat{\cal L}V\right)
   \nonumber \\
   &-& 2\mu\hat\Omega\left(\hat{\cal L}W\right)
   - 2\left(1-\mu^2\right)\frac{\partial\left(\mu\hat\Omega\right)}
   {\partial\mu}\frac{\partial W}{\partial\mu}
   -2 m^2\frac{\partial\hat\Omega}{\partial\mu} W
   \nonumber \\
   &+& 2\mu\hat\Omega_\mathrm{A}\left(\hat{\cal L}B\right)
   + 2\left(1-\mu^2\right)\frac{\partial
   \left(\mu\hat\Omega_\mathrm{A}\right)}
   {\partial\mu}\frac{\partial B}{\partial\mu}
   +2 m^2\frac{\partial\hat\Omega_\mathrm{A}}{\partial\mu} B
   \nonumber \\
   &-& m\hat\Omega_\mathrm{A}\left(\hat{\cal L}A\right)
   - 2m\frac{\partial\left(\mu\hat\Omega_\mathrm{A}\right)}
   {\partial\mu} A
   - 2m\left(1-\mu^2\right)\frac{\partial\hat\Omega_\mathrm{A}}
   {\partial\mu}
   \frac{\partial A}{\partial\mu}
   \nonumber \\
   &+& m\hat\Omega\left(\hat{\cal L}V\right)
   + 2m\frac{\partial\left(\mu\hat\Omega\right)}{\partial\mu} V
   + 2m\left(1-\mu^2\right)\frac{\partial\hat\Omega}{\partial\mu}
   \frac{\partial V}{\partial\mu},
   \label{11}
\end{eqnarray}
where $\mu = \cos\theta$, and
\begin{equation}
   \hat\lambda = \frac{N}{\Omega_0 k r}
   \label{12}
\end{equation}
can be understood as special normalization for radial wavelength.
The first term on the righthand side describes the stabilizing
effect of the stratification. It vanishes for small $\hat\lambda$.
Apart from this  stabilizing buoyancy term, the wavelength is only
present in diffusive terms. The second term of the RHS  includes
the action of  finite viscosity,
\begin{equation}
  \epsilon_\nu = \frac{\nu N^2}{\Omega_0^3 r^2}.
\label{13}
\end{equation}
Similarly, we use below
\begin{equation}
  \epsilon_\eta = \frac{\eta N^2}{\Omega_0^3 r^2},\ \ \ \ \ \ \
  \epsilon_\chi = \frac{\chi N^2}{\Omega_0^3 r^2}
  \label{13a}
\end{equation}
for the diffusive parameters $\eta$ and $\chi$. The second and the
following lines of Eq.~(\ref{11}) describe the influences of the
basic rotation and  the toroidal field. Only latitudinal
derivatives of $\Omega$ and $\Omega_\mathrm{A}$ appear. All radial
derivatives are absorbed by disturbances which vary on much
shorter radial scales than $\Omega$ or $\Omega_\mathrm{A}$. The
complete system of five equations also includes the equation for
toroidal flow
\begin{eqnarray}
  \hat\omega\left(\hat{\cal L}W\right) &=&
  - \mathrm{i}\frac{\epsilon_\nu}{\hat\lambda^2}\left(\hat{\cal L}W\right)
  + m\hat\Omega\left(\hat{\cal L}W\right)
  - m\hat\Omega_\mathrm{A}\left(\hat{\cal L}B\right)
  \nonumber \\
  &-& mW\frac{\partial^2}{\partial\mu^2}
  \left(\left(1-\mu^2\right)\hat\Omega\right)
  + mB\frac{\partial^2}{\partial\mu^2}
  \left(\left(1-\mu^2\right)\hat\Omega_\mathrm{A}\right)
  \nonumber \\
  &+& \left(\hat{\cal L}V\right)\frac{\partial}{\partial\mu}
  \left(\left(1-\mu^2\right)\hat\Omega\right)
  - \left(\hat{\cal L}A\right)\frac{\partial}{\partial\mu}
  \left(\left(1-\mu^2\right)\hat\Omega_\mathrm{A}\right)
  \nonumber \\
  &+& \left(\frac{\partial}{\partial\mu}\left(\left(1-\mu^2\right)^2
  \frac{\partial\hat\Omega}{\partial\mu}\right) -
  2\left(1-\mu^2\right)\hat\Omega\right)\frac{\partial V}{\partial\mu}
  \nonumber \\
  &-& \left(\frac{\partial}{\partial\mu}\left(\left(1-\mu^2\right)^2
  \frac{\partial\hat\Omega_\mathrm{A}}{\partial\mu}\right) -
  2\left(1-\mu^2\right)\hat\Omega_\mathrm{A}\right)
  \frac{\partial A}{\partial\mu},
  \label{14}
\end{eqnarray}
the equation for toroidal magnetic field
\begin{eqnarray}
   \hat\omega\left(\hat{\cal L}B\right)&=&
   - \mathrm{i}\frac{\epsilon_\eta}{\hat\lambda^2}\left(\hat{\cal L}B\right)
   + m\hat{\cal L}\left(\hat\Omega B\right)
   - m\hat{\cal L}\left(\hat\Omega_\mathrm{A} W\right)
   \nonumber \\
   &-& m^2\frac{\partial\hat\Omega}{\partial\mu} A
   - \frac{\partial}{\partial\mu}\left(
   \left(1-\mu^2\right)^2\frac{\partial\hat\Omega}{\partial\mu}
   \frac{\partial A}{\partial\mu}\right)
   \nonumber \\
   &+& m^2\frac{\partial\hat\Omega_\mathrm{A}}{\partial\mu} V
   + \frac{\partial}{\partial\mu}\left(
   \left(1-\mu^2\right)^2\frac{\partial\hat\Omega_\mathrm{A}}{\partial\mu}
   \frac{\partial V}{\partial\mu}\right) ,
   \label{15}
\end{eqnarray}
the poloidal field equation
\begin{equation}
   \hat\omega\left(\hat{\cal L}A\right) =
   - \mathrm{i}\frac{\epsilon_\eta}{\hat\lambda^2}\left(\hat{\cal L}A\right)
   + m\hat\Omega\left(\hat{\cal L}A\right)
   - m\hat\Omega_\mathrm{A}\left(\hat{\cal L}V\right)
   \label{16}
\end{equation}
and the entropy equation,
\begin{equation}
   \hat\omega S = - \mathrm{i}\frac{\epsilon_\chi}{\hat\lambda^2}
   S + m\hat\Omega S + \hat{\cal L}V .
   \label{17}
\end{equation}

In the simplest case of $\Omega=\Omega_\mathrm{A}=0$ and vanishing
diffusivities the only nontrivial solution of the above equations
are short (in radius) gravity waves with
\begin{equation}
   \omega = \frac{N}{kr}\sqrt{l(l+1)},\ \ l=1,2,...
   \label{18}
\end{equation}
The instabilities we shall find are thus due to either magnetic
field or nonuniform rotation. It should be kept in mind that the
above equations are only valid for $kr \gg 1$. We shall see that
the maximum growth rates do indeed belong to the short radial
scales.
\subsection{2D approximation ($\hat\lambda\gg 1$)}\label{2D}
Generally, the ratio of $N/\Omega_0$ in radiative zones is  so
large (Fig.~\ref{f1}) that $\hat{\lambda}$ can also be large in
spite of $kr \gg 1$. Equation (\ref{11}) in the leading order in
$\hat\lambda$ then gives $S = 0$. Then Eqs.~(\ref{17}) and
(\ref{16}) successively yield $V=0$ and $A=0$. Diffusive terms can
also be neglected. The equation system reduces to two coupled
equations for toroidal magnetic field and toroidal flow (Gilman \&
Fox \cite{GF97}),
\begin{eqnarray}
  \hat\omega\left(\hat{\cal L}W\right) &=&
  m\hat\Omega\left(\hat{\cal L}W\right)
  - m\hat\Omega_\mathrm{A}\left(\hat{\cal L}B\right)
  \nonumber \\
  &-& mW\frac{\partial^2}{\partial\mu^2}
  \left(\left(1-\mu^2\right)\hat\Omega\right)
  + mB\frac{\partial^2}{\partial\mu^2}
  \left(\left(1-\mu^2\right)\hat\Omega_\mathrm{A}\right),
  \nonumber \\
  \hat\omega\left(\hat{\cal L}B\right)&=&
   m\hat{\cal L}\left(\hat\Omega B\right)
   - m\hat{\cal L}\left(\hat\Omega_\mathrm{A} W\right).
   \label{19}
\end{eqnarray}
Here  the wave number $k$ drops out and the equations describe 2D
disturbances within decoupled spherical shells.

The particular case where $\Omega$ and $\Omega_\mathrm{A}$ are
both constant simplifies equations (\ref{19}) strongly. The
equations have constant coefficients in this case and can be
solved analytically. The solution in terms of Legendre polynomials
$W,B\sim P^m_l(\mu )$ leads to the eigenfrequencies
\begin{equation}
   \frac{\hat\omega}{m} = 1 - \frac{1}{l(l+1)} \pm
   \sqrt{\hat\Omega_\mathrm{A}^2\left( 1 - \frac{2}{l(l+1)}\right)
   + \frac{1}{l^2(1+l)^2}   }
   \label{21}
\end{equation}
of magnetically modified $r$-modes (Longuet-Higgins \cite{LH64};
Papaloizou \& Pringle \cite{PP78}) describing stable horizontal
patterns drifting in longitude. We shall see in Section \ref{one
belt} that the case of constant $\Omega$ and $\Omega_\mathrm{A}$
shows Tayler instability with 3D approach. 2D approximation,
however, misses the instability. Though the result is obtained for
particular case of constant $\Omega_\mathrm{A}$ it is most
probably valid in general. 2D incomressive distortions on
spherical surfaces do not change area encircled by toroidal
magnetic field lines. The circular lines of background field have
minimum length for given encircled area. Any distortion increases
the length of closed field lines thus increasing magnetic energy.
There is no possibility to feed an instability by magnetic energy
release. Only with differential rotation 2D instabilities are
possible (Gilman \& Fox \cite{GF97}).

\section{Results and discussion}
We proceed by discussing numerical solutions of the perturbation
equations for special profiles of $\Omega_\mathrm{A}$. In the
present paper the interaction of toroidal magnetic fields and the
basic rotation is formulated only for rigid rotation. The work
with  differential rotation is much more complicated, but it is in
progress. In the Appendix as a first announcement the {\em
hydrodynamic} stability of latitudinal differential rotation in 3D
is discussed.

For the toroidal field two simple geometries are considered.
First, the quantity  $\Omega_\mathrm{A}$ is taken constant so that
the toroidal field has only one belt symmetric with respect to the
equator. Second, two magnetic belts are considered with equatorial
antisymmetry, i.e. with a node of $B_\phi$ at the equator.

\subsection{Fields with equatorial symmetry}\label{one belt}
Constant $\Omega$ and $\Omega_\mathrm{A}$ give the simplest
realization of the Tayler instability. The instability criterion
for nonaxisymmetric disturbances (Goossens et al. \cite{GBT81})
reads
\begin{equation}
   \frac{\partial}{\partial\mu}\ln\left(B_\phi^2\right) <
   \frac{2\mu^2 - m^2}{\mu\left(1-\mu^2\right)} ,\ \ \
   \mathrm{for}\ \mu\geq 0.
   \label{20}
\end{equation}
It is satisfied with constant $\Omega_\mathrm{A}$ for $m=1$ and
close to the poles.

From Section 2.3 we know that the instability can appear only in
3D formulation when radial displacements are allowed. For this
case the equations (\ref{11}), (\ref{14})--(\ref{17}) have been
solved numerically.
\begin{figure}[htb]
   \centering
   \centering
   \includegraphics[width=7.0cm]{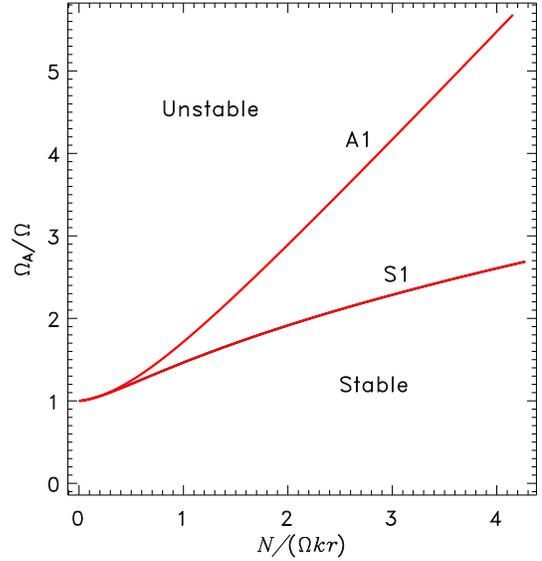}
   \caption{The stability map for constant $\Omega_\mathrm{A}$
            and zero diffusivities. There is no instability for
            weak fields with $\Omega_\mathrm{A} < \Omega$.
            The threshold field strength for the instability increases
            with increasing vertical wavelength $\hat\lambda$.
              }
   \label{f2}
\end{figure}

Consider first ideal fluids with zero diffusivities, $\chi = \eta
= \nu = 0$. Figure ~\ref{f2} gives the resulting  stability map.
Standard notations, Sm and Am, are used for the symmetry types of
the global modes with azimuthal wave number $m$ symmetric and
antisymmetric to the equator. However, the disturbances of
different fields of the same mode possess different symmetries.
Rather arbitrarily, our symmetry notations correspond to the
symmetry of the potential $W$ of toroidal field. We did not find
instabilities for $m\neq 1$.

The smallest field strength producing instability corresponds to
shortest radial scales, $\hat\lambda \rightarrow 0$. The radial
velocity is also zero in this limit. Note that putting $u'_r = 0$
regardless of which value the vertical wavelength, $\lambda$, has
returns to the 2D case without any instability. The point here is
that for  $u'_r = 0$ the entire poloidal flow vanishes. The
horizontal part of the flow is proportional to $u'_r r/\lambda$. A
horizontal poloidal flow is necessary for the pinch-type
instability and the flow remains finite when radial velocity and
radial scale both approach zero keeping their ratio constant. The
largest growth rates correspond to this limit. In this sense, the
instability indeed proceeds via \lq\emph{almost} horizontal'
displacements. The tendency for the instability to prefer
indefinitely short radial scales shows that finite diffusivities
should be included.

Figure~\ref{f2} does not show any instability for weak fields with
$\Omega_\mathrm{A}<\Omega$. With other words, the basic  rotation
makes a strong stabilizing effect. This may be a special property
of our model  where the ratio $\Omega_\mathrm{A}/\Omega$  is
uniform (Pitts \& Tayler \cite{PT85}). We indeed find that the
threshold field for $\Omega_\mathrm{A}\sim\cos\theta$ is about ten
times smaller compared to constant $\Omega_\mathrm{A}$. A much
stronger destabilization, however, is produced by finite
diffusion.

From now on  the values
\begin{equation}
 \epsilon_\nu = 2\cdot 10^{-10}, \ \ \ \ \ \ \ \
 \epsilon_\eta = 4\cdot 10^{-8}, \ \ \ \ \ \ \ \
 \epsilon_\chi = 10^{-4},
   \label{22}
\end{equation}
which are characteristic for the upper part of the solar radiative
core, are used for the diffusion parameters (\ref{13}). The
relations $\chi\gg\eta\gg\nu$ of Eq.~(\ref{22}) are quite typical
of stellar radiation zones.
\begin{figure}[htb]
   \centering
   \centering
   \includegraphics[width=7.0cm]{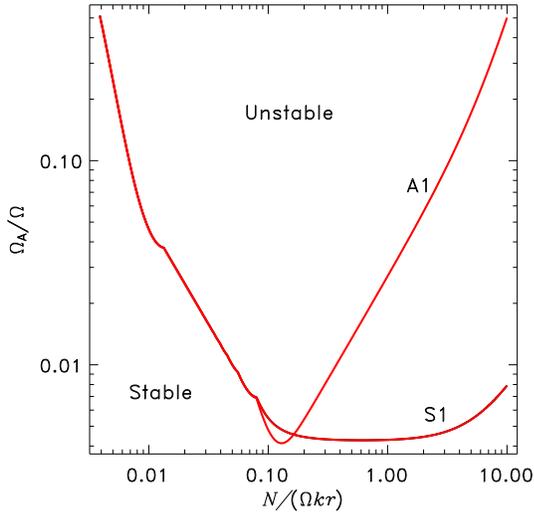}
   \caption{The same as in Fig. \ref{f2} but with the  finite
            diffusivities (\ref{22}). The critical field
            strengths for the onset of the instability are strongly reduced
            compared to ideal MHD.
              }
   \label{f3}
\end{figure}

For ideal fluids the Tayler instability operates with extremely
small radial scales. When there is no stabilizing stratification,
however, finite vertical scales are preferred (Arlt et al.
\cite{ASR07}). The thermal conductivity decreases the stabilizing
effect of the stratification and reduces strongly the critical
field strengths for the instability. The characteristic minima in
Fig.~\ref{f3} correspond to small, $\hat\lambda \la 1$, but finite
vertical scales.
\begin{figure}
   \centering
   \includegraphics[width=7.0cm]{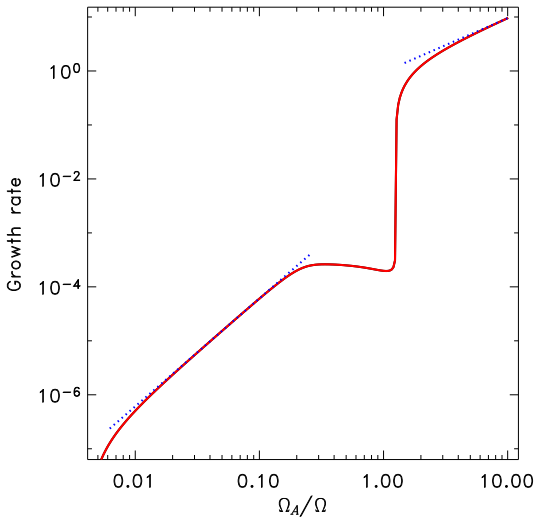}
   \caption{The normalized growth rate $\hat\sigma$
            as function of the magnetic field
            amplitude  for the  S1 mode and for
        $\hat\lambda = 0.6$ (where the line of Fig.~\ref{f3}
        has a minimum). The dotted lines for strong,
        $\Omega_\mathrm{A} > \Omega$, and weak,
        $\Omega_\mathrm{A} < \Omega$, fields give  approximations
        by the power laws $\hat\sigma \sim \hat\Omega_\mathrm{A}$ and
            $\hat\sigma \sim \hat\Omega^2_\mathrm{A}$ respectively.
              }
   \label{f4}
\end{figure}

The growth rates for weak fields ($\Omega_\mathrm{A} < \Omega$)
where the instability exists due to finite diffusion, are very
small. For strong fields ($\Omega_\mathrm{A} > \Omega$) the basic
rotation is not important and  $\Omega_\mathrm{A}$ scales the
growth rate. The dependence of Fig.~\ref{f4} does indeed approach
the $\sigma \sim \Omega_\mathrm{A}$ relation ($\sigma$ is the
growth rate) in the strong-field limit. The growth rate drops by
almost four orders of magnitude when $\hat\Omega_\mathrm{A}$ is
reduced below 1. In the weak-field regime it is  $\sigma \sim
\Omega^2_\mathrm{A}/\Omega$ (Spruit \cite{S99}).
\begin{figure}[htb]
   \centering
   \includegraphics[width=8.0cm]{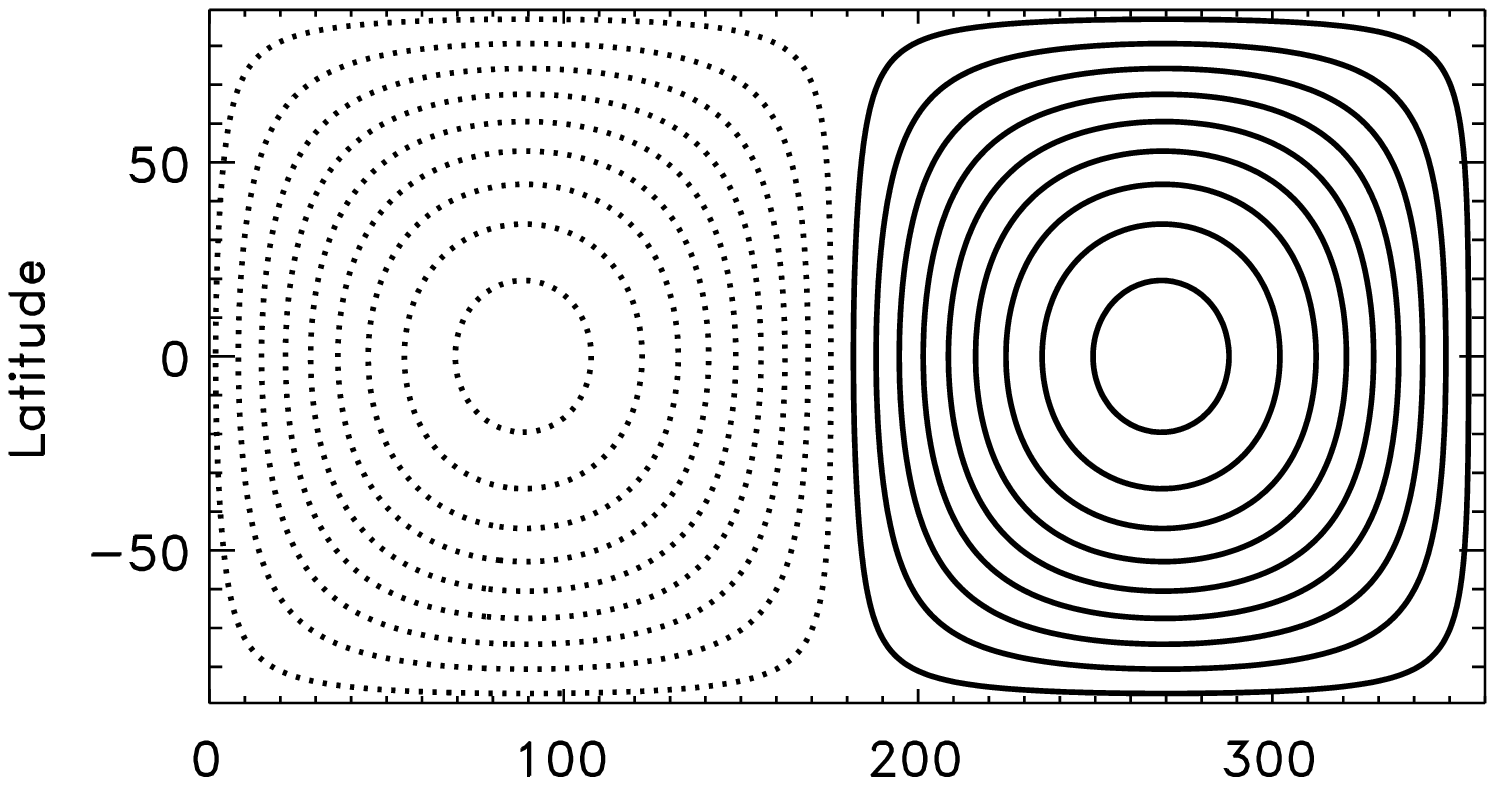}
   \includegraphics[width=8.0cm]{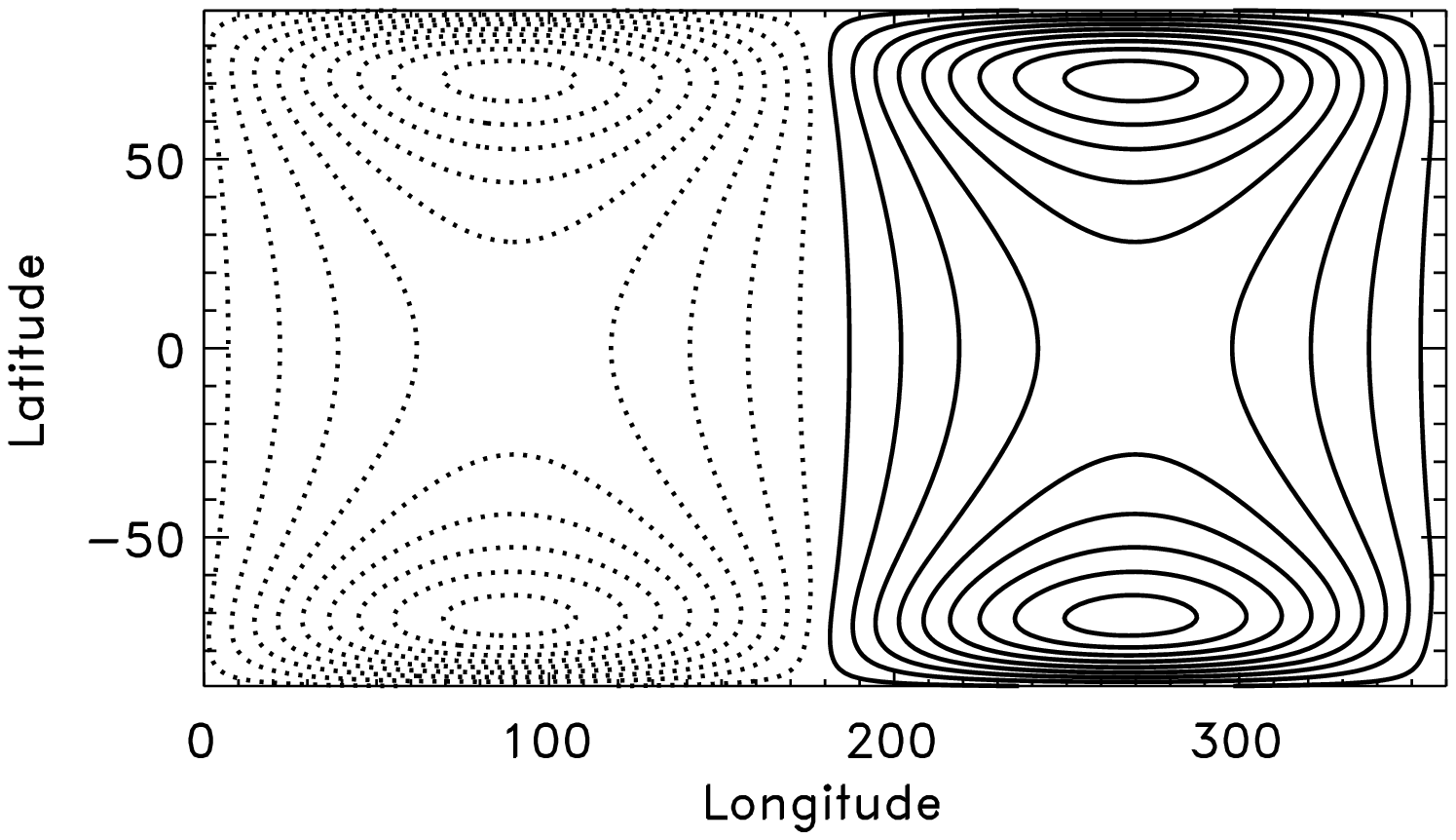}
   \caption{Toroidal field lines of the most rapidly growing eigenmodes
            of Fig.~\ref{f4} for weak background toroidal field,
        $\Omega_\mathrm{A} = 0.1\Omega$ (upper panel), and strong
        field, $\Omega_\mathrm{A} = 10\Omega$ (lower panel). The horizontal axis gives the longitude.
              }
   \label{f5}
\end{figure}

The structures of the unstable modes differ between strong and
weak field regimes. For the weak fields the resulting pattern is
distributed over the entire sphere. For strong fields it is much
more concentrated to the poles (Fig~\ref{f5}) but it remains
global in latitude.
\subsection{Fields with equatorial antisymmetry}
\begin{figure}[htb]
   \centering
   \includegraphics[width=7.0cm]{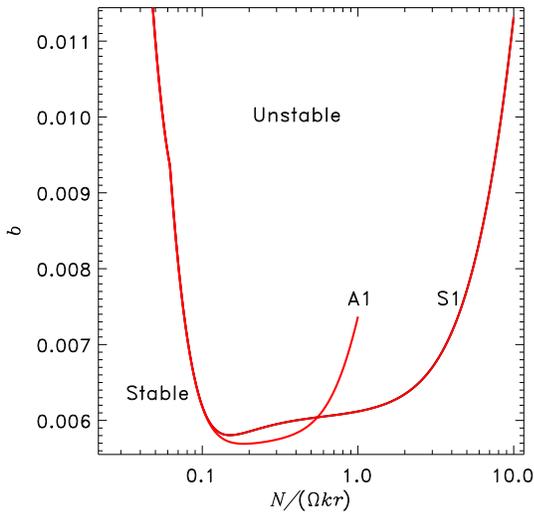}
   \caption{Stability map for the  field model (\ref{23}) with  two
             belts and equatorial antisymmetry.
              }
   \label{f6}
\end{figure}

Now a simple toroidal field with two belts of opposite polarity,
i.e.
\begin{equation}
   \hat\Omega_\mathrm{A} = b\cos\theta,
   \label{23}
\end{equation}
is considered with the diffusivity set (\ref{22}). Such a field
geometry can result from the action of differential rotation on
dipolar poloidal fields. Figure~\ref{f6} shows the corresponding
stability map. It is similar to the map of Fig.~\ref{f3}.
Instability is again found only for $m=1$.

\begin{figure}[htb]
   \centering{
   \includegraphics[width=4.1cm]{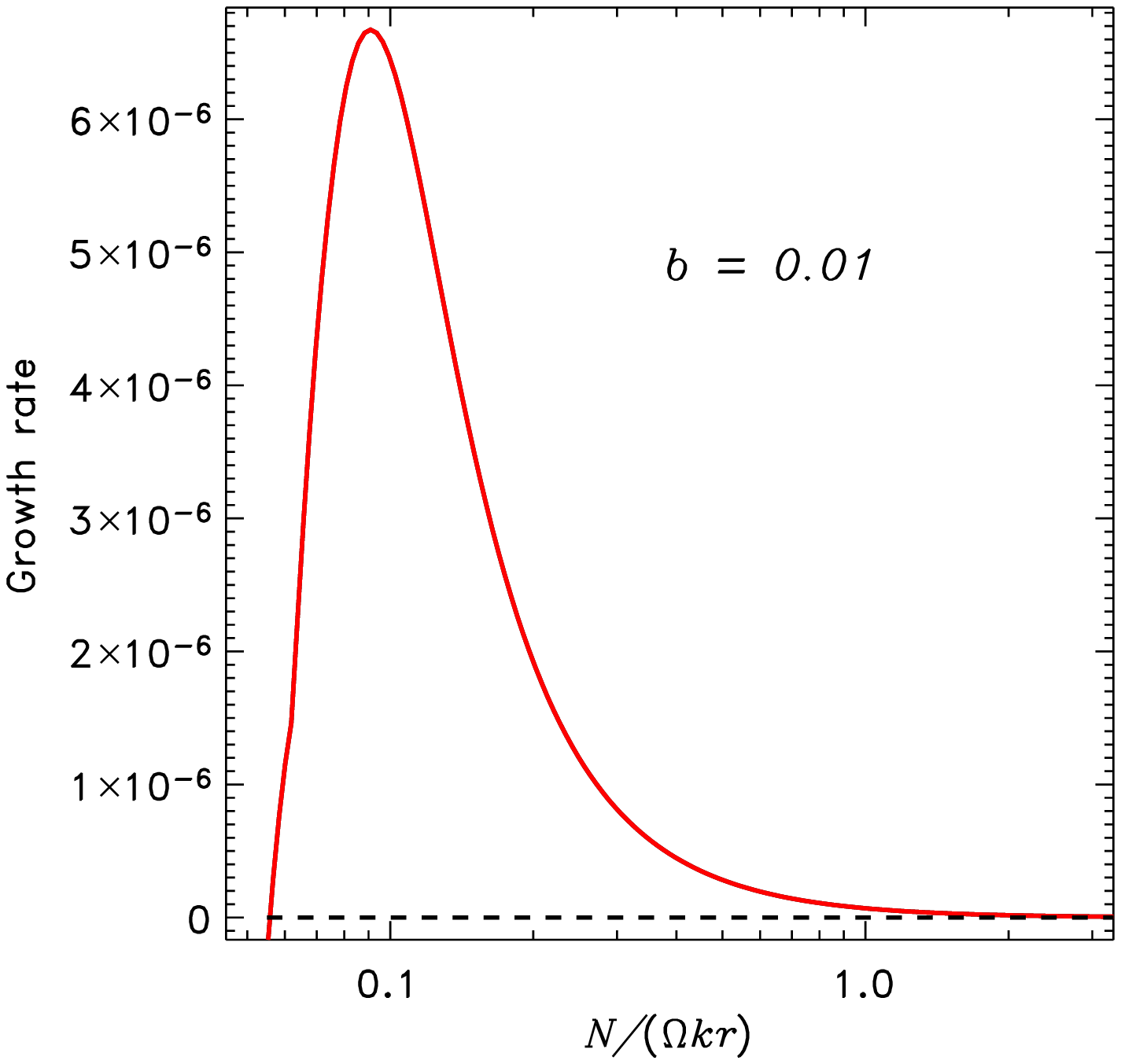}
   \hspace{0.2cm}
   \includegraphics[width=4.3cm]{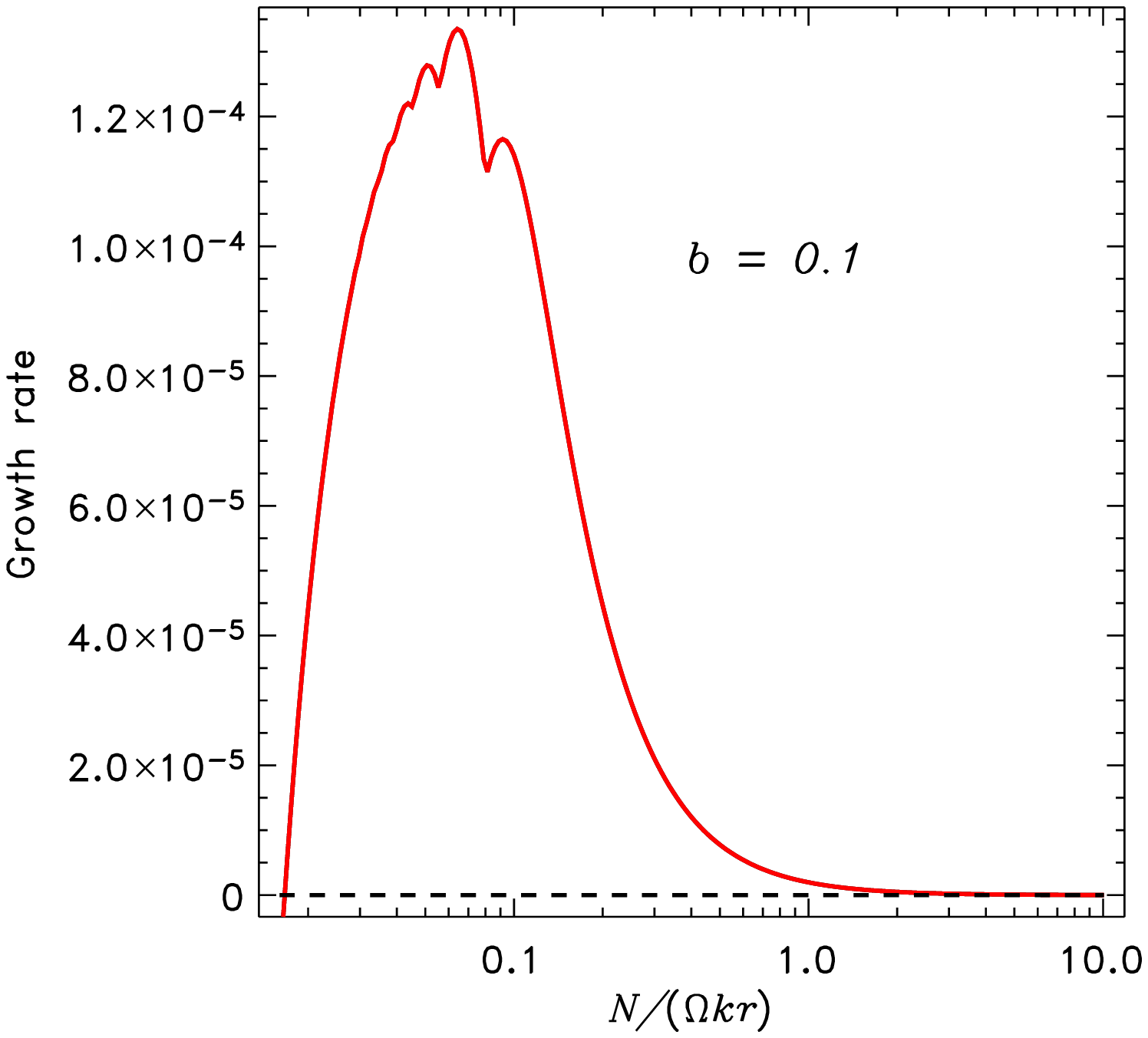}}
   \caption{Growth rates in units of $\Omega$  for $b=0.01$ (left) and $b=0.1$ (right).
        The largest rates exist for   $\hat\lambda \;\buildrel
        <\over{\scriptstyle\sim}\; 0.1$ independent of the magnetic field
        amplitude. All for S1 modes.
              }
   \label{f7}
\end{figure}

In physical values for the upper part of solar radiation zone
(with $\rho\simeq$ 0.2 g/cm$^3$) one finds
\begin{equation}
    B_\phi \simeq 10^5 b \ \ \ \ \mathrm{G}
    \label{24}
\end{equation}
and
\begin{equation}
     \lambda \simeq 10\hat\lambda\ \mathrm{Mm},
     \label{25}
\end{equation}
so that  from  Fig.~\ref{f6} follows a critical magnetic field for
the instability slightly below 600~G. The mode which first becomes
unstable if  the field exceeds  this critical  value has a
vertical wavelength between 1 and 2 Mm. For larger field
strengths, of course, there is a range of unstable wavelengths.
The maximum growth rates appears, however, at wavelengths
\lower.4ex\hbox{$\;\buildrel <\over{\scriptstyle\sim}\;$}1~Mm
(Fig.~\ref{f7}).

\begin{figure}[htb]
   \centering
   \includegraphics[width=7.0cm]{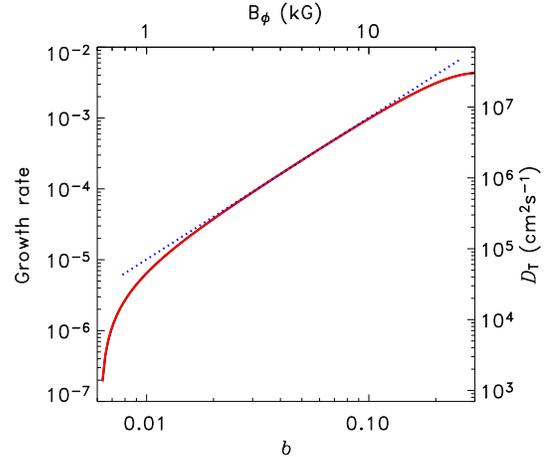}
   \caption{The growth rate as function of the toroidal field
            amplitude for $\hat{\lambda} = 0.1$. 
            The dotted line shows the parabolic approximation
            $\hat\sigma = 0.1 b^2$. The scale
            on the right gives the estimated radial diffusivity
            of chemical species after (\ref{26}).
              }
   \label{f8}
\end{figure}

The flow field of any  instability mixes chemical species also in
radial direction. Such an  instability can thus  be relevant to
the radial transport of the light elements (Barnes et al.
\cite{BCM99}). The effective diffusivity, $D_\mathrm{T} \simeq
u'\ell$ ($u'$ and $\ell$ are rms velocity and correlation length
in radial direction) can  roughly be estimated from our linear
computations assuming that $\sigma \simeq \ell/u'$ and $\ell
\simeq \lambda/2$. With Eq.\~(\ref{25}) this yields
\begin{equation}
    D_\mathrm{T} \simeq 7 \cdot 10^9\ \hat\sigma\ \
    \mathrm{cm}^2\mathrm{s}^{-1} ,
    \label{26}
\end{equation}
where $\hat\sigma$ is the normalized growth rate given in
Fig.~\ref{f8}. For the range of $0.01 < b <0.2$ where the plot is
closely approximated by the parabolic law $\hat\sigma \simeq
0.1b^2$, Eq. (\ref{26}) can be rewritten in terms of $B_\phi$
(Eq.~(\ref{24})) as
\begin{equation}
     D_\mathrm{T} \simeq 7\cdot 10^4 \left(\frac{B_\phi}{1\ \mathrm{kG}}
     \right)^2\ \ \ \ \mathrm{cm}^2\mathrm{s}^{-1}.
     \label{27}
\end{equation}

As known, diffusivities  exceeding    $10^3$\,cm$^2$s$^{-1}$ in
the upper radiative core are not compatible with the observed
solar lithium abundance. Hence, the toroidal field amplitude can
only slightly exceed the marginal value of about 600~G. In our
(simplified) formulation, the observed solar lithium abundance
seem to {\em exclude}\footnote{or rise new problems} any concept
of hydromagnetic dynamos driven by Tayler instability in the upper
radiation zone of the Sun. We should not forget, however, that the
superrotation ($\partial \Omega/\partial r >0$) at the bottom of
the convection zone in the equatorial region acts {\em
stabilizing} so that the critical field amplitutes for Tayler
instability may be higher than the computed 600 G.

\section{Summary}
Linear stability of toroidal magnetic field in rotating stellar
radiation zones is analyzed assuming that the  vertical scale of
the fluctuations is short compared to the  local radius
(`short-wave approximation'). The analysis is global in horizontal
dimensions. Stability computations confirm that the most rapidly
growing perturbations have short radial scales: $kr \sim 10^3$.

We have shown that pinch-type instability of toroidal field
require nonvanishing radial displacements. The instability does
not appear in 2D approximation with zero radial velocities. The
maximum amplitude of stable toroidal magnetic fields for the Sun
which we have found is about $600$~G. This value results only for
rigid rotation. It will most probably increase if the stabilizing
influence of the positive radial gradient of $\Omega$ in the
equatorial region of the tachocline is included into the model.

The field strength  in the upper part of the solar radiative
interior can only marginally exceed the resulting  critical
values. Otherwise the instability would produce too strong radial
mixing of light elements. After our results all the axisymmetric
hydromagnetic models of the solar tachocline (R\"udiger \&
Kitchatinov \cite{RK97}; Garaud \cite{G02}; Sule et al.
\cite{SRA05}; Kitchatinov \& R\"udiger \cite{KR06}) have stable
toroidal fields. On the contrary, the strong fields $\sim 10^5$~G
which are able to modify $g$-modes or even stronger fields  which
may influence solar neutrinos are strongly unstable with e-folding
times shorter than one rotation period.

3D computations of joint instabilities of toroidal fields and
differential rotation (Gilman \& Fox \cite{GF97}; Cally
\cite{C03}, R\"udiger et al. 2007) can be a perspective for
further work. Another tempting extension is the inclusion of the
poloidal field. The field can be important in view of the very
short vertical scales of the unstable modes.

In the Appendix we present a calculation with the same equations
for the hydrodynamic instability of latitududinal differential
rotation. This instability can already be found in 2D
approximations (Watson 1981) but it is substantially modified in
the 3D theory.

\begin{acknowledgements}
This work was supported by the Deutsche Forschungsgemeinschaft
and by the Russian Foundation for Basic Research (project 05-02-04015).
\end{acknowledgements}

\begin{appendix}
\section{The Watson problem in 3D}\label{Watson}
Latitudinal differential rotation can be unstable even without
magnetic field if the shear  $\partial\Omega/\partial \theta$ is
sufficiently large (Watson \cite{W81}; Dziembowski \& Kosovichev
\cite{DK87}). The instability may reduce the differential rotation
to its critical  value (Garaud \cite{G01}) which  can be relevant
to the theory of the solar tachocline. The critical relative value
of 28\% for differential rotation found by Watson resulted from a
2D theory (cf. Section \ref{2D}). The value has also appeared in a
3D numerical probe of marginal stability of a shell rotating fast
enough with the rotation law
\begin{equation}
    \Omega = \Omega_0\left( 1 - a\cos^2\theta\right),
    \label{28}
\end{equation}
but for not stratified material (Arlt, Sule \& R\"udiger 2007).
The critical shear increases to much higher values, however,  if
the real rotation law (including its radial variation) of the
solar tachocline is adopted.

Equations (\ref{11}), (\ref{14}), and (\ref{17}) of the Section
\ref{equations} (in their hydrodynamical version) can also be
applied to extend the Watson approach by allowance for radial
displacements. Note that the Reynolds number ${\rm Re} = \Omega
r^2/\nu$ can be written as
\begin{equation}
{\rm Re} = \frac{N^2/\Omega^2}{\epsilon_\nu},
\label{Re}
\end{equation}
which with Eq.~(\ref{22}) gives a very large value, O(10$^{15}$).

\begin{figure}[htb]
   \centering
   \includegraphics[width=7.0cm]{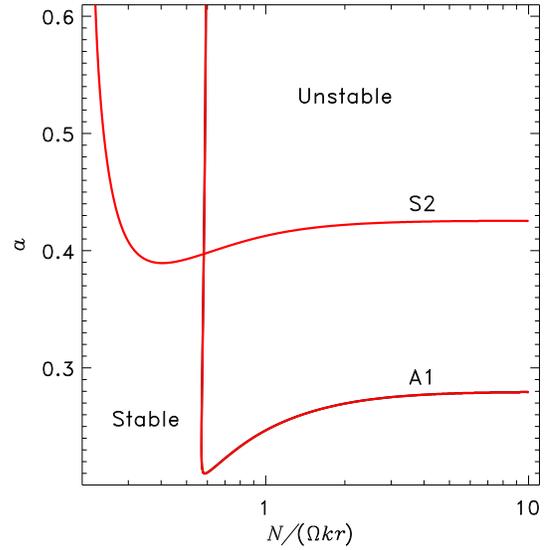}
   \caption{Stability map for the hydrodynamic instability of
            latitudinal differential rotation. Most unstable are the
            perturbations of A1 symmetry type with the vertical scale
            $\hat\lambda \simeq 0.6$. The critical magnitude of latitudinal
            shear is reduced to 0.21 compared to the 0.28 value of 2D
            theory.
              }
   \label{f9}
\end{figure}

For positive and sufficiently large $a$, the modes A1 and S2
become unstable. Figure~\ref{f9} shows the dependence of the
critical values of $a$ on the normalized wavelength $\hat\lambda$
(\ref{12}). For large enough radial wavelengths the 28\%-value of
the 2D theory is reproduced. It is reduced, however, to $a = 0.21$
in 3D calculations. We see that short rather than long radial
scales are preferred. The minimum $a$ appears for $\hat\lambda
\simeq 0.6$, so that the characteristic wavelength of $\lambda
\simeq 6$~Mm results after Eq.~(\ref{25}) for the solar
tachocline.

\end{appendix}
\end{document}